\documentclass[twocolumn,showpacs,preprintnumbers,amsmath,amssymb]{revtex4}
\usepackage{graphics,graphicx}% Include figure files
\usepackage{dcolumn}% Align table columns on decimal point
\usepackage{bm}% bold math
\begin{document}
\bibliographystyle{prsty}
\newcommand{\bq}{\begin{mathletters}}
\newcommand{\eq}{\end{mathletters}}
\newcommand{\beq}{\begin{eqnarray}}
\newcommand{\eeq}{\end{eqnarray}}
\newcommand{\beqq}{\begin{eqnarray*}}
\newcommand{\eeqq}{\end{eqnarray*}}

\newcommand{\st}[1]{{\mbox{${\mbox{\scriptsize #1}}$}}}  % subscript or superscript text
\newcommand{\BM}[1]{\mbox{\boldmath $#1$}}              % boldmath
\newcommand{\uv}[1]{\mbox{$\widehat{\mbox{\boldmath $#1$}}$}}   % unit vector
\newcommand{\dy}[1]{\mbox{\boldmath $\overline{#1}$}}   % dyadic
\newcommand{\sbtex}[1]{{\mbox{\scriptsize #1}}}   % subscript/superscript sized text
\newcommand{\nab}{\mbox{\boldmath $\nabla$}}            % bold nabla
\newcommand{\CS}{\mbox{$\begin{array}{c}\cos \\ \sin \\ \end{array}$}}
\newcommand{\SC}{\mbox{$\begin{array}{c}\sin \\ \cos \\ \end{array}$}}

\newcommand{\rmi}{{\rm i}}
\newcommand{\er}{{\bf e}_{r}}
\newcommand{\ep}{{\bf e}_\varphi}
\newcommand{\et}{{\bf e}_\theta}
\newcommand{\hh}{{\bf H}}
\newcommand{\ee}{{\bf E}}
\newcommand{\ww}{{\bf W}}
\newcommand{\vv}{{\bf v}}
\newcommand{\rr}{{\bf r}}
\newcommand{\eee}{{\bf e}}
\newcommand{\uu}{{\bf u}}
\newcommand{\lom}{{\bf L}}

\newcommand{\rmq}{{\rm q}}
\newcommand{\rmt}{{\rm t}}
\newcommand{\co}{{\rm co}}
\newcommand{\cl}{{\rm cl}}

\newcommand{\epr}{\varepsilon_{r r}}
\newcommand{\mur}{\mu_{r r}}
\newcommand{\alr}{\alpha_{r r}}
\newcommand{\kar}{\kappa_{r r}}
\newcommand{\Om}{\Omega_a^r}
\newcommand{\ve}{\vec{e}}
\newcommand{\vw}{\vec{w}}
\newcommand{\va}{\vec{a}}

%\preprint{submitted to Physical Review E}

\title{ Inverse problem in transformation optics }
\author{A.V. Novitsky$^{1,2,*}$}

\affiliation{$^1$DTU Fotonik, Department of Photonics Engineering,
Technical University of Denmark, ${\O}$rsteds~plads~343, \\ DK-2800 Kgs. Lyngby, Denmark
\\
$^2$Department of Theoretical Physics, Belarusian State University,
Nezavisimosti~Avenue~4, 220030 Minsk, Belarus \\
$^*$Corresponding author: anov@fotonik.dtu.dk}

\date{\today}

\begin{abstract}
The straightforward method of transformation optics implies that one starts from the coordinate transformation, determines the Jacobian matrix, the fields and material parameters of the cloak. However, the coordinate transformation appears as an optional function: it is not necessary to know it. We offer the solution of some sort of the inverse problem: starting from the fields in the invisibility cloak we directly derive the permittivity and permeability tensors of the cloaking shell. The approach can be useful for finding material parameters for the specified electromagnetic fields in the cloaking shell without knowing coordinate transformation.
\end{abstract}

\maketitle

\section{Introduction}

Precursor theoretical papers \cite{Dolin, Ward, Pendry03,Alu} and the rapid development of the metamaterials as technological branch of optics and material physics had led to the discovery of the exciting transformation optics \cite{Pendry}. The most interesting application of the transformation optics is obviously the invisibility cloaking. Some groups recently reported on the realization of the experimental invisibility of large objects \cite{Chen,Zhang} using the carpet cloaking \cite{Li,Ma}. Transformation optics approach is extended now to describe the field concentrators \cite{Rahm}, perfect lensing \cite{Leonhardt}, illusion optics \cite{LaiCM,LaiIO,LiExp}, optics in curved space \cite{Schultheiss}, optical analogues of the cosmological redshift \cite{Ginis}, Aharonov-Bohm effect and black hole horizon \cite{Leonhardt}.

In general, transformation optics is an approach dealing with the change of material parameters and field distributions owing to the transformation of a spatial region. The transformation is usually a sort of compression or tension. It can be described by the coordinate transformation converting electromagnetic (original) space to the physical (transformed) space. Since the objects to hide are commonly in the free space, the original space is the vacuum. The squeeze of vacuum of electromagnetic space corresponds to the shell in physical space. Spatial region inside the shell does not interact with the incident radiation. Therefore, it is not visible and can be filled with any material or contain any object inside. The cloaking shell is characterized by the very complex material parameters: (i) both dielectric permittivity and magnetic permeability tensors should be equal, and (ii) material should be anisotropic and inhomogeneous. Nevertheless the invisibility cloaking has been already experimentally verified \cite{Schurig,Cai,Gabrielli,Alitalo,Ma}.

There are several methods to derive the expressions of transformation optics. Except the classical \cite{Pendry} and general-relativity-based \cite{Leonhardt} approaches the alternative methods can be mentioned. Primarily it is the approach of Tretyakov et al \cite{Tretyakov}, which introduces the general concept of field transformations defined as linear relation between the original and transformed fields. The transformation optics of Yaghjian et al \cite{Yaghjian} uses the field boundary conditions, but not the boundary conditions embedded into the coordinate transformation as usual. The so called method of generating functions (or inverse approach) \cite{Novitsky,Qiu} is offered  for some special cases of spherical and cylindrical cloaks. The generating function (for instance, some component of the permittivity tensor) is used instead of the coordinate transformation.

The approach of this paper is not intended for the replacement of the ordinary method of transformation optics. The aim of the approach is to find the cloaking parameters of the material for the predefined fields. The reconstruction of the medium parameters from the field distribution can be called an inverse problem in transformation optics by analogy with the inverse problem in scattering theory. On the one hand, we use the known results of transformation optics, such as the formulae for the fields and material parameters expressed in terms of the Jacobian matrix. On the other hand, the coordinate transformation in explicit form is not necessary. In the approach, we specify the fields inside the cloak assuming the incident plane wave. Then the dielectric permittivity and magnetic permeability tensors are derived using the three scalar field potentials describing the wave phase, electric field vector, and magnetic field vector.

\section{Inverse problem}

Let the electric and magnetic fields in the electromagnetic and physical spaces are specified. We are looking for the Jacobian matrix $J$ and material parameters $\varepsilon$ and $\mu$ corresponding to such fields. Electromagnetic fields in electromagnetic $\ee' (\rr')$, $\hh'(\rr')$ and physical $\ee (\rr)$, $\hh(\rr)$ spaces are assumed to be functions of the coordinates of their own spaces, that is $\rr' = (x'_1, x'_2, x'_3)$ and $\rr = (x_1, x_2, x_3)$, respectively. In general, the Jacobian matrix is the function of coordinates in both electromagnetic and physical spaces, because $J(\rr,\rr')$ connects the coordinates $\rr$ and $\rr'$. The link between the coordinates (coordinate transformation) can be found using the equation $\partial x_j / \partial x'_i = J_{ij}$. Then the Jacobian matrix can be written as a quantity of the single space, $J(\rr)$ or $J(\rr')$. In this paper we are not interested in the coordinate transformation. Below we will make sure that the coordinate transformation is not necessary for finding dielectric permittivity and magnetic permeability of the transformed medium.

\begin{figure}
\centerline{ \includegraphics[scale=0.6,clip=]{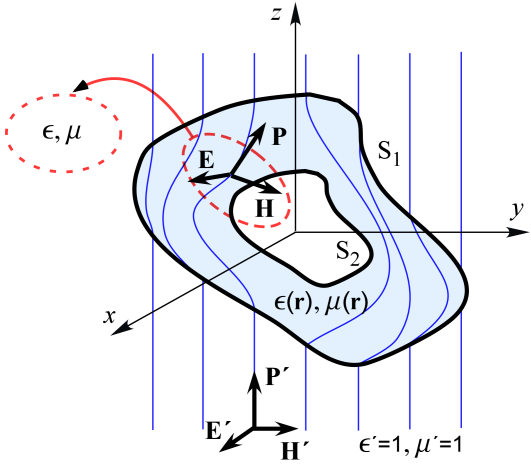} }
\caption{ Sketch demonstrating the geometry of the problem. } \label{fig:1}
\end{figure}

In electromagnetic space, the plane wave propagates in $z'$-direction in vacuum $\varepsilon'=1$ and $\mu'=1$. Fields of the plane wave are described in the common way: $\ee' = \exp(\rmi k_0 z') \eee_x$ and $\hh' = \exp(\rmi k_0 z') \eee_y$ (see figure \ref{fig:1}), where $k_0 = \omega/c$ is the wavenumber in vacuum, $\omega$ is the circular frequency, and $c$ is the speed of light in vacuum. So, we consider the transformation from the plane wave in vacuum to the complex field distribution in physical space. Since an arbitrary wave can be presented as the superposition of plane waves, we expect that the invisibility cloaking for the plane wave will give the material parameters applicable for any field distribution in the electromagnetic space.

%%%%%%%%%%%%%%%%%%%%%%%%%%%%%%%%%

%%%%%%%%%%%%%%%%%%%%%%%%%%%%%%%%%

Boundary conditions are usually specified via the coordinate transformation: the spatial region in electromagnetic space confined by the surface $S_1$ is squeezed to the region between the surfaces $S_1$ and $S_2$. If the coordinate transformation is not defined, the boundaries of the transformed space should be set by means of the fields. We have four ``no-scattering'' conditions (the continuity of the electric and magnetic fields tangential to the outer interface $S_1$ defining the impedance matching with the ambient medium) and the ``no-field-inside-cloak'' condition, which can be formulated as zero energy flux passing through the inner boundary (field does not penetrate):
\begin{eqnarray}
&& {\bf n}_1 \times \ee(S_1) = {\bf n}_1 \times \ee'(S_1), \quad {\bf n}_1 \times \hh(S_1) = {\bf n}_1 \times \hh'(S_1), \nonumber \\
&& {\bf n}_2 {\bf S}(S_2) \equiv \frac{c}{4 \pi} {\bf n}_2 [\ee(S_2) \times \hh(S_2)] = 0, \label{BC_fields}
\end{eqnarray}
where ${\bf n}_1$ and ${\bf n}_2$ are the unit vectors normal to the surfaces $S_1$ and $S_2$, respectively. For example, the interfaces $S_1$ and $S_2$ can be spheres or cylinders. But more complicated geometries are possible, too. We emphasize that only the inner and outer interfaces are specified by equation (\ref{BC_fields}), but not the coordinate transformation.

Further we will find the inverse Jacobian matrix from equations of transformation optics
\begin{equation}
\ee(\rr) = J^{-1}(\rr,\rr') \ee'(\rr'), \qquad \hh(\rr) = J^{-1}(\rr,\rr') \hh'(\rr'). \label{eJehJh}
\end{equation}
%Three-dimensional matrix $J$ has 9 elements to be found. Since we have only 6 equations (\ref{eJehJh}), we can definitely determine only 6 components. Three elements (a 3D vector) stay free.

Solution of the first vector equation can be written as the sum of the contributions along the electric field $\ee'$ and orthogonal to the electric field $\ee'$:
\begin{equation}
J^{-1} = \frac{\ee \otimes \ee'}{\ee^{\prime 2}} + \hat{\alpha} \ee^{\prime \times}, \label{J1}
\end{equation}
where $\hat{\alpha}$ is an arbitrary matrix. Dyad $\ee \otimes \ee'$ (in index form $(\ee \otimes \ee')_{ij} = E_i E'_j$) operates as a projector for arbitrary vector ${\bf a}$: $(\ee \otimes \ee'){\bf a} = \ee (\ee'{\bf a})$ and ${\bf a}(\ee \otimes \ee') = ({\bf a}\ee) \ee'$. Three-dimensional antisymmetric tensor $\ee^{\prime \times}$ dual to the vector $\ee'$ \cite{Fedorov,Fedorov58} can be represented in index form as $(\ee^{\prime \times})_{ik} = \sum_{j=1}^3 \varepsilon_{ijk} E'_j$, where $\varepsilon_{ijk}$ is the Levi-Civita symbol. Dual tensor is characterized by the following properties: $\ee^{\prime \times} {\bf a} = \ee^{\prime} \times {\bf a}$ and ${\bf a} \ee^{\prime \times} = {\bf a} \times \ee^{\prime}$. By substituting solution (\ref{J1}) into the second equation in (\ref{eJehJh}) we obtain the equation with respect to matrix $\hat{\alpha}$:
\begin{equation}
\hh - \frac{(\ee' \hh')}{\ee^{\prime 2}} \ee = \hat{\alpha} (\ee' \times \hh').
\end{equation}
The solution of this equation can be presented in the similar form as (\ref{J1}). Using the relationship ${\bf P}^{\prime \times} \ee^{\prime \times} = \ee' \otimes {\bf P}'$ we finally derive the Jacobian matrix
\begin{equation}
J^{-1}(\rr,\rr') = \frac{\ee \otimes \hh' - \hh \otimes \ee'}{{\bf P}^{\prime 2}} {\bf P}^{\prime \times} + {\bf a} \otimes {\bf P}', \label{Jfinal}
\end{equation}
where quantity ${\bf P}' = \ee' \times \hh'$ is proportional to the Poynting vector ${\bf S}'$ of the plane wave in electromagnetic space. An arbitrary vector ${\bf a}$ (three elements of matrix $J$) cannot be found from equation (\ref{eJehJh}).

Actually we have found more than we expected, because from expression (\ref{Jfinal}) it follows that the Jacobian matrix remains the same, if the fields in electromagnetic and physical spaces are simultaneously multiplied by the same scalar function $f'(\rr') = f(\rr)$. This means that the Jacobian matrix is invariant with respect to the phase factor. If $\ee' = \exp(\rmi k_0 z') \eee_x$ and $\hh' = \exp(\rmi k_0 z') \eee_y$, then the fields in physical space are $\ee = \exp(\rmi k_0 z'(\rr)) J^{-1} \eee_x$ and $\hh = \exp(\rmi k_0 z'(\rr)) J^{-1} \eee_y$. Therefore, we can specify just fields $\ee' = \eee_x$ and $\hh' = \eee_y$ ignoring the same scalar functions, which are terminated in the definition of Jacobian matrix. So, we derive
\begin{equation}
J^{-1}(\rr) = \ee_0 \otimes \ee'_0 + \hh_0 \otimes \hh'_0 + {\bf a}_0 \otimes {\bf P}'_0, \label{Jfinal1}
\end{equation}
where $\ee'_0 = \eee_x$, $\hh'_0 = \eee_y$, ${\bf P}'_0 = \eee_z$, $\ee_0 = \exp(-\rmi k_0 z') \ee$, $\hh'_0 = \exp(-\rmi k_0 z') \hh$, and ${\bf a}_0 = \exp(2\rmi k_0 z') {\bf a}$.

Vectors $\ee_0$, $\hh_0$, and ${\bf a}_0$ of the physical space are not arbitrary. In addition to the boundary conditions (\ref{BC_fields}) restricting the electric and magnetic fields, the Jacobian matrix $J$ is not arbitrary as well. It should be agreed with Maxwell's equations or, in terms of the transformation optics, the identity condition
\begin{equation}
\nabla \times J^{-1}(\rr) = 0 \label{rotJ0}
\end{equation}
should hold. Introducing matrix (\ref{Jfinal1}) to equation (\ref{rotJ0}) we get to tree vector relationships
\begin{equation}
\nabla \times \ee_0 = 0, \qquad \nabla \times \hh_0 = 0, \qquad \nabla \times {\bf a}_0 = 0. \label{rotEHa}
\end{equation}
Potential vectors $\ee_0$, $\hh_0$, and ${\bf a}_0$ defined by (\ref{rotEHa}) can be presented by means of the scalar potentials $\psi$, $\chi$, and $\eta$ as
\begin{equation}
\ee_0 = \nabla \psi, \qquad \hh_0 = \nabla \chi, \qquad {\bf a}_0 = \nabla \eta,
\end{equation}
respectively. The meaning of the potentials can be determined from the comparison of the Jacobian matrix
\begin{equation}
J^{-1} = \nabla \otimes (\psi \eee_x + \chi \eee_y + \eta \eee_z)
\end{equation}
with its definition $J^{-1} = \nabla \otimes \rr'$. The correspondence is clear: scalar potentials are the coordinates in electromagnetic space, that is $x' = \psi(\rr)$, $y' = \chi(\rr)$, and $z' = \eta(\rr)$. These three equations define the coordinate transformation. Since the plane wave in electromagnetic space propagates in the $z'$-direction, the phase factor is described by the potential $\eta$ as $\exp(\rmi k_0 z') = \exp(\rmi k_0 \eta)$. The potentials have clear physical meaning for the fields: $\psi$ and $\chi$ define the vector structure of the electric and magnetic fields, while $\eta$ is the phase function. Thus, the electric and magnetic fields in the cloaking shell are
\begin{equation}
\ee = {\rm e}^{\rmi k_0 \eta(\rr)} \nabla \psi(\rr), \qquad \hh = {\rm e}^{\rmi k_0 \eta(\rr)} \nabla \chi(\rr). \label{FieldsViaPotent}
\end{equation}
where $\psi$, $\chi$, and $\eta$ are the electric, magnetic, and phase potentials, respectively.

Dielectric permittivity and magnetic permeability tensors equal
\begin{eqnarray}
\varepsilon = \mu = \frac{J^T J}{\sqrt{\det (J^T J)}} = \sqrt{\det(\hat{A})} \hat{A}^{-1}, \nonumber \\
\hat{A}(\rr) = \nabla \psi \otimes \nabla \psi + \nabla \chi \otimes \nabla \chi + \nabla \eta \otimes \nabla \eta,
\label{MatParametersFromEH}
\end{eqnarray}
where the superscript T denotes the transpose operation. It is important that the permittivity and permeability tensors are defined in the physical space. That is why the coordinate transformation is not required and the tensors are set in terms of the potentials --- field characteristics. It should be mentioned that we do not differ the covariant and contravariant indices in three-dimensional tensors. We use the tensors in index-free form, e.g. $\varepsilon = (\varepsilon_{ij}) = (\varepsilon^{ij})$. The methods of operating with index-free objects are well developed in Refs. \cite{Fedorov,Fedorov58,BarkovskyFurs} and applied in the optics of complex (anisotropic, bianisotropic) media \cite{Barkovskii,Borzdov,Novitsky05,QiuNovitsky}.

One more limitation on the potentials is caused by the fact that the Jacobian matrix is nonsingular, i.e. $\det(J^{-1}) \neq 0$ or $\det(\hat{A}) \neq 0$. In other words, the vectors $\ee_0$, $\hh_0$, and ${\bf a}_0$ are not parallel: $(\ee_0 \times \hh_0){\bf a}_0 \neq 0$. Since $\ee_0$ and $\hh_0$ are not parallel, vector ${\bf a}_0$ should surely have the component along ${\bf P}_0 = \ee_0 \times \hh_0$.

In general, all potentials are important for setting electric and magnetic fields in the cloaking shell. However, if only vectorial distributions of the electric and magnetic fields are essential for us, the phase of the field can be omitted. Then the same vectorial distributions follow for various phase potentials and we can tune the phase potential to find the simplest form of the permittivity and permeability tensors.

How should the potentials look like? We can suppose that they should be like Cartesian coordinates written in curvilinear coordinates, because the meaning of the potentials are the Cartesian coordinates in the electromagnetic space. For example, potentials $\psi = x' = R(r) \sin\theta \cos\varphi$, $\chi = y' = R(r) \sin\theta \sin\varphi$, and $\eta = z' = R(r) \cos\theta$ can be applied for determining material tensors in spherical coordinates. Detailed discussion of the possible forms of the potentials is given in the subsequent sections.

\section{Spherical cloak}

Let us consider a spherical cloak with inner and outer boundaries $a$ and $b$, respectively. External electric and magnetic fields (they coincide with the fields in the electromagnetic space)
\begin{eqnarray}
\ee' (\rr) = {\rm e}^{\rmi k_0 r \cos\theta} \left( \sin\theta \cos\varphi \eee_r + \cos\theta \cos\varphi \eee_\theta - \sin\varphi \eee_\varphi \right), \nonumber \\
\hh' (\rr) = {\rm e}^{\rmi k_0 r \cos\theta} \left( \sin\theta \sin\varphi \eee_r + \cos\theta \sin\varphi \eee_\theta + \cos\varphi \eee_\varphi \right) \nonumber
\end{eqnarray}
are linked with electric and magnetic fields inside the cloak
\begin{eqnarray}
\ee (\rr) = {\rm e}^{\rmi k_0 \eta(\rr)} \left( \frac{\partial \psi}{\partial r} \eee_r + \frac{1}{r} \frac{\partial \psi}{\partial \theta} \eee_\theta + \frac{1}{r \sin\theta} \frac{\partial \psi}{\partial \varphi} \eee_\varphi \right), \nonumber \\
\hh (\rr) = {\rm e}^{\rmi k_0 \eta(\rr)} \left( \frac{\partial \chi}{\partial r} \eee_r + \frac{1}{r} \frac{\partial \chi}{\partial \theta} \eee_\theta + \frac{1}{r \sin\theta} \frac{\partial \chi}{\partial \varphi} \eee_\varphi \right) \label{fields_SC}
\end{eqnarray}
via the conditions of continuity of the tangential fields (\ref{BC_fields}) at the outer boundary $r=b$.
These conditions can be written in the form
\begin{eqnarray}
\frac{\partial \psi(b,\theta,\varphi)}{\partial \theta} = b {\rm e}^{\rmi k (b \cos\theta - \eta(b,\theta,\varphi))} \cos\theta \cos\varphi, \nonumber \\
\frac{\partial \psi(b,\theta,\varphi)}{\partial \varphi} = - b {\rm e}^{\rmi k (b \cos\theta - \eta(b,\theta,\varphi))} \sin\theta \sin\varphi, \nonumber \\
\frac{\partial \chi(b,\theta,\varphi)}{\partial \theta} = b {\rm e}^{\rmi k (b \cos\theta - \eta(b,\theta,\varphi))} \cos\theta \sin\varphi, \nonumber \\
\frac{\partial \chi(b,\theta,\varphi)}{\partial \varphi} = b {\rm e}^{\rmi k (b \cos\theta - \eta(b,\theta,\varphi))} \sin\theta \cos\varphi.
\end{eqnarray}
For the agreement of the equations a couple of partial differential equations $\partial^2 \psi/\partial\theta \partial\varphi = \partial^2 \psi/\partial\varphi \partial\theta$ and $\partial^2 \chi/\partial\theta \partial\varphi = \partial^2 \chi/\partial\varphi \partial\theta$ with respect to the function $\eta(b,\theta,\varphi)$ should hold true. In the current situation, this couple of equations is easily solved giving $\eta(b,\theta,\varphi) = b \cos\theta + \eta_0$. Electric and magnetic potentials at $r = b$ follow from the integration: $\psi(b,\theta,\varphi) = b \exp(-\rmi k \eta_0) \sin\theta \cos\varphi + \psi_0$ and $\chi(b,\theta,\varphi) = b \exp(-\rmi k \eta_0) \sin\theta \sin\varphi + \chi_0$, where $\eta_0$, $\psi_0$, and $\chi_0$ are constants. It can be an arbitrary dependence of the potential on the radial coordinate $r$ reducing to $\psi(b,\theta,\varphi)$, $\chi(b,\theta,\varphi)$, and $\eta(b,\theta,\varphi)$ at $r=b$. From the great variety of functions we choose
\begin{eqnarray}
\psi(r,\theta,\varphi) &=& f(r) \sin\theta \cos\varphi, \quad \chi(r,\theta,\varphi) = g(r) \sin\theta \sin\varphi, \nonumber \\
\eta(r,\theta,\varphi) &=& h(r) \cos\theta,
\end{eqnarray}
where $f$, $g$, and $h$ are arbitrary functions except fulfilment of the following boundary conditions: $f(b) = g(b) = h(b) = b$, and $f(a) = 0$ or $g(a) = 0$. Two last conditions follow from the vanishing of the Poynting vector component normal to the inner sphere $r = a$: $S_r \sim f(a) g(a) = 0$. The cloak is realized for any $f$, $g$, and $h$ discussed above. So, we can choose these functions to provide the needed electric and magnetic field vectors (and phase, if required). Then the permittivity and permeability tensors are easily calculated from equation (\ref{MatParametersFromEH}).

If the phase is not important, potential $\eta$ can be used for optimization of the permittivity and permeability tensors. Assuming $f(r) = g(r) \neq h(r)$ we derive
\begin{eqnarray}
\varepsilon = \frac{f P}{r^2 T} \eee_r \otimes \eee_r + \frac{f P}{T} \eee_\theta \otimes \eee_\theta + \frac{T}{f} \eee_\varphi \otimes \eee_\varphi \nonumber \\
+ \frac{f \cos\theta (h' h - f' f)}{r T} \left(\eee_r \otimes \eee_\varphi + \eee_\varphi \otimes \eee_r \right),
\end{eqnarray}
where $P(r,\theta) = f^2 \cos^2\theta + h^2 \sin^2\theta$, $T(r,\theta) = h' f \cos^2\theta + f' h \sin^2\theta$, $f' = df/dr$. The same distribution of the electric $|\ee|^2$ and magnetic $|\hh|^2$ fields and time-averaged Poynting vector $(c/8\pi) {\rm Re}(\ee \times \hh^\ast)$ is realized for any phase function $h(r)$. Generally dielectric permittivity tensor depends on two coordinates, $r$ and $\theta$, and has non-diagonal elements. The simple phase functions $h = b$ or $h = r$ do not eliminate the $\theta$-dependence or anisotropy. They can be eliminated only for the equal radial functions, i.e. $f(r) = g(r) = h(r)$. Then the permittivity tensor takes the well-known form
\begin{equation}
\varepsilon = \frac{f^2}{r^2 f'} \eee_r \otimes \eee_r + f' \eee_\theta \otimes \eee_\theta + f' \eee_\varphi \otimes \eee_\varphi. \label{parameters_sph}
\end{equation}

%%%%%%%%%%%%%%%%%%%%%%%%%%%%%%%%%

\begin{figure}
\centerline{ \includegraphics[scale=0.6,clip=]{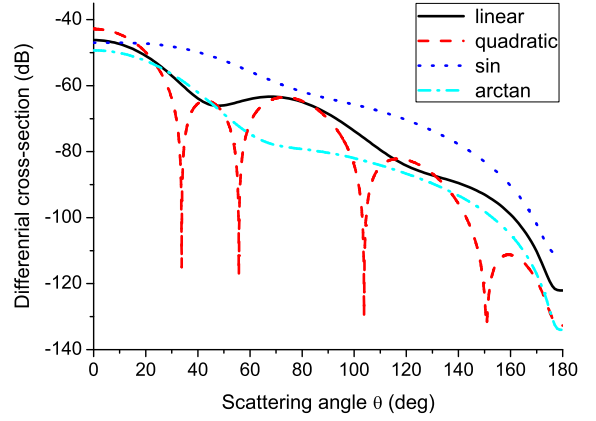} }
\caption{ Differential cross-sections for spherical cloaks with different field profiles $f(r)$: linear $f(r) = b (r-a)/(b-a)$, quadratic $f(r) = b (r^2-a^2)/(b^2-a^2)$, sinusoidal $f(r) = b \sin[(\pi/2) (r-a)/(b-a)]$, and arctan-form $f(r) = (2 b/\pi) \arctan[(r-a)/(b-r)]$. Parameters: $k_0 a = \pi$, $k_0 b = 2 \pi$. } \label{fig:2}
\end{figure}

%%%%%%%%%%%%%%%%%%%%%%%%%%%%%%%%%

%%%%%%%%%%%%%%%%%%%%%%%%%%%%%%%%%

\begin{figure}
\centerline{ \includegraphics[scale=0.6,clip=]{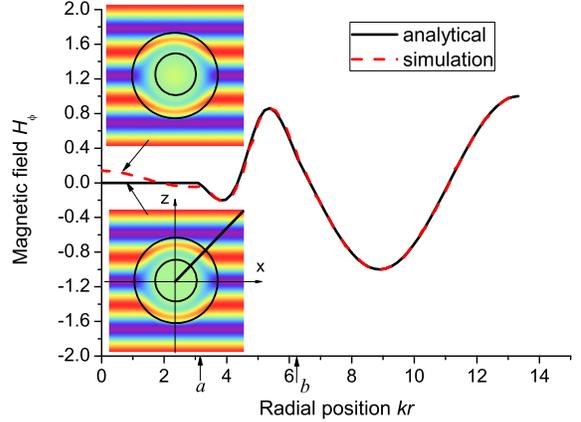} }
\caption{ Magnetic field component $H_\varphi$ of the arctan-form cloak for $\varphi = 0$ and $\theta = 45^\circ$ calculated both analytically and using the simulation. In the insets the magnetic field distributions in the plane $\varphi = 0$ are shown. Parameters: $k_0 a = \pi$, $k_0 b = 2 \pi$. } \label{fig:3}
\end{figure}

%%%%%%%%%%%%%%%%%%%%%%%%%%%%%%%%%

Example of the more sophisticated choice of the potentials is
\begin{eqnarray}
\psi(r,\theta,\varphi) = f(r) \sin\Theta(r,\theta) \cos\Phi(r,\varphi) + \psi_0(r), \nonumber \\
\chi(r,\theta,\varphi) = g(r) \sin\Theta(r,\theta) \sin\Phi(r,\varphi) + \chi_0(r), \nonumber \\
\eta(r,\theta,\varphi) = h(r) \cos\Theta(r,\theta),
\end{eqnarray}
where $f(b) = g(b) = h(b) = b$, $\Theta(b,\theta) = \theta$, and $\Phi(b,\varphi) = \varphi$.

Any function $f(r) = g(r) = h(r)$ meeting the appropriate boundary conditions ($f(b) = b$ and $f(a) = 0$) can be used to obtain the cloaking parameters (\ref{parameters_sph}). In figure \ref{fig:2}, the differential cross-sections for four field profiles $f(r)$ are shown. Inhomogeneous cloaking shell is divided into $N=30$ homogeneous spherical layers. Differential cross-section normalized by the geometrical cross-section of the spherical particle is calculated using the analytical expressions given in Ref. \cite{QiuNovitsky}. As it has been expected, all curves in figure \ref{fig:2} show good cloaking properties in the whole angle range.

In figure \ref{fig:3}, we compare two solutions for the cloak with arctan profile: one curve is obtained as the closed-form solution with $f(r) = (2 b/\pi) \arctan[(r-a)/(b-r)]$, another curve is computed for the appropriate cloaking parameters using the field scattering technique. The fields in the cloak are very close one to another. The only region, where they differ, is the region inside the cloaking shell. It is explained by the inaccuracy of the simulation which cannot precisely evaluate the fields near $r=a$ (the hidden region is filled with glass).

\section{Cylindrical cloak}

%%%%%%%%%%%%%%%%%%%%%%%%%%%%%%%%%

\begin{figure}
\centerline{ \includegraphics[scale=0.6,clip=]{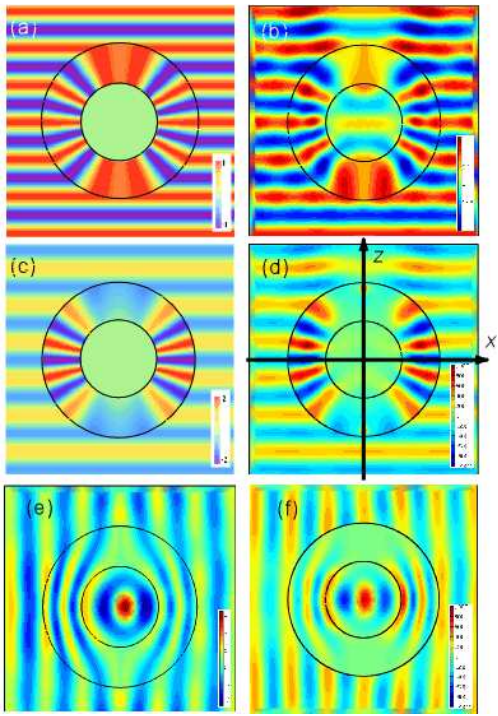} }
\caption{ Magnetic fields $H_y$ evaluated (a) analytically and (b) in simulation. Electric field component $E_x$ obtained (c) analytically and (d) using the simulation. (e) Magnetic $H_y$ and (f) electric $E_z$ fields simulated in the case of the plane wave incident in $x$ direction. Parameters: $k_0 a = \pi$, $k_0 b = 2 \pi$. } \label{fig:4}
\end{figure}

%%%%%%%%%%%%%%%%%%%%%%%%%%%%%%%%%

The similar algorithm can be applied for the cylindrical cloak extended from $r=a$ to $r=b$ (cylinder axis is $y$-directed).

(i) Write down the electric and magnetic fields in and outside the cloak. Fields in the cloaking shell are given by equation (\ref{FieldsViaPotent}).

(ii) Apply the boundary conditions (\ref{BC_fields}) at the outer $S_1$ and inner $S_2$ interfaces.
%: $\psi(b,\varphi,y) = b \cos\varphi \exp(-\rmi k \eta_0) + \psi_0$, $\chi(b,\varphi,y) = y \exp(-\rmi k \eta_0) + \chi_0$, and $\eta(b,\varphi,y) = b \sin\varphi \exp(-\rmi k \eta_0) + \eta_0$.

(iii) Choose the needed form of the potentials $\psi$, $\chi$, and $\eta$ and determine the fields (\ref{FieldsViaPotent}).

(iv) Calculate the permittivities using equation (\ref{MatParametersFromEH}).

We pick out the potentials for the cylindrical cloak in the form
\begin{eqnarray}
\psi(r,\varphi,z) = f(r) \cos\varphi, \qquad \chi(r,\varphi,z) = g(r) y, \nonumber \\
 \eta(r,\varphi,z) = h(r) \sin\varphi.
\end{eqnarray}
Then the fields are equal to
\begin{eqnarray}
\ee = {\rm e}^{i k_0 h(r) \sin\varphi} \left( \frac{d f}{d r} \cos\varphi \eee_r - \frac{f(r)}{r} \sin\varphi \eee_\varphi \right), \nonumber \\
\hh = {\rm e}^{i k_0 h(r) \sin\varphi} \left( \frac{d g}{d r} y \eee_r + g(r) \eee_y \right),
\end{eqnarray}
where $f(b) = h(b) = b$, $g(b) = 1$ and $f(a) = 0$ or $g(a) = 0$. The material parameters of the ordinary cylindrical cloak can be obtained for $f(r) = h(r)$ and $g(r) = 1$.
%Cylindrical cloak has one infinite dimension. Therefore, the additional conditions on the fields $\ee(\rr \rightarrow \infty) \neq \infty$ and $\hh(\rr \rightarrow \infty) \neq \infty$ should be put %on. So, we conclude that $g(r) = 1$ and $g(a) = 0$ cannot be agreed.

If $f \neq h$, we can expect more complicated material tensors. Magnetic field in the cloak equals $\hh = \exp(\rmi k \eta) \eee_y$. It possesses the unit amplitude and can differ only in phase. We take $f(r) = b(r-a)/(b-a)$ and $h(r) = b$ satisfying the required boundary conditions. Function $f$ shall be chosen to provide the needed electric field. The phase potential in our case is very simple: $\eta = b \sin\varphi$. Nevertheless, the permittivity tensor is not diagonal in the basis of the cylindrical coordinates:
\begin{equation}
\varepsilon = \mu = \left( \begin{array}{ccc} \varepsilon_{rr} & \varepsilon_{r\varphi} & 0 \\ \varepsilon_{r\varphi} & \varepsilon_{\varphi\varphi} & 0 \\ 0 & 0 & \varepsilon_{yy} \end{array} \right), \label{eps_cyl}
\end{equation}
where
$\varepsilon_{rr} = [(b-a)^2 + (r-a)^2 \tan\varphi]/[(b-a) r]$, $\varepsilon_{\varphi\varphi} = r/(b-a)$, $\varepsilon_{yy} = b^2 \cos^2\varphi]/[(b-a) r]$, and $\varepsilon_{r\varphi} = (r-a) \tan\varphi/(b-a)$. In figure \ref{fig:4}(a) and (c) we demonstrate the fields, which we expect in the cloak. They are plotted using the closed-form expressions $H_y = \exp(\rmi k b \sin\varphi)$ and $E_x = \exp(\rmi k b \sin\varphi) (1 - a \sin^2\varphi/r)b/(b-a)$, respectively. To verify that we will obtain the similar field distributions in the cloaking shell with parameters (\ref{eps_cyl}), the numerical simulation was carried out. From figures \ref{fig:4}(b) and (d) it is obvious that the cloaking shell indeed guides the prescribed fields. However, it will be another field distribution, if the incident wave propagates along $x$ axis (or in some direction differing from $z$ axis). In figures \ref{fig:4} (e) and (f) exactly such a situation is on view. One indicates the cloaking property of the shell in this case, too.

\section{Concluding remarks}

The inverse approach discussed in this paper can be applied to other transformation optics devices. For example, concentrator consists of the inner and outer regions \cite{Rahm,Jiang}. The inner region follows from the squeeze of the electromagnetic space, while the outer one is the result of extension. Both transformed spatial regions are characterized by the parameters (\ref{FieldsViaPotent}), but with their own potentials: $\psi_1$, $\chi_1$, $\eta_1$ and $\psi_2$, $\chi_2$, $\eta_2$ for outer and inner regions, respectively. The potentials are not arbitrary. They are restricted by the boundary conditions put on the fields. The impedance matching conditions are the first two vector equation in (\ref{BC_fields}). Two more equations are the continuity of the tangential components at the interface between the inner and outer regions. The field will be concentrated in the inner region, if there is a squeeze, i.e. condition $\det(J^{-1})>1$ holds true.

The second example for application is the cylindrical-to-plane-wave convertor \cite{JiangConv}. The idea of this device is to modify the cylindrically symmetric region to the part of a square. Then the cylindrical electromagnetic wave generated at the center will turn to the plane wave after transmission through the transformed region. In our notations, there is a part of square with the plane wave electromagnetic field outside. The fields inside the square have the form (\ref{FieldsViaPotent}). We put only the impedance-matching boundary conditions (\ref{BC_fields}) on the fields to obtain the convertor.

In conclusion, we have solved the inverse problem in transformation optics, that is we derived the analytical formula (\ref{MatParametersFromEH}) for the material tensors using the information about the fields inside the cloaking shell: the phase and vector distributions of the electric and magnetic fields. The closed-form expressions are not general. They are written for the incident plane wave and $\varepsilon' = \mu' = 1$ in the electromagnetic space. Nevertheless, there is no limitations on the form of the cloak. Moreover, the derived material tensors describe the cloak for any incident wave, not only for the plane wave used to obtain $\varepsilon$ and $\mu$. If any other wave is incident, the cloaking property of the shell keeps, but the field distribution in the cloak is not specified now. It is well known that material parameters of a wide class of cylindrical cloaks have a singularity at the inner boundary. We believe that the inverse approach can be useful in finding nonsingular parameters of the cloaking shell. The example of nonsingular parameters is equation (\ref{eps_cyl}). The method developed seems to be very attractive as the means of deeper understanding of the physics and mechanisms of the optical cloaking and can substantially help in design of the cloaks with predefined fields inside.

Financial support from the Danish Research Council for Technology and Production Sciences via project THz COW and Basic Research Foundation of
Belarus (grant F10M-021) is acknowledged.

\end{document}